\documentclass[twocolumn,english,amsmath,amssymb,superscriptaddress,nofootinbib]{revtex4-1}
\usepackage[T1]{fontenc}
\usepackage[latin9]{inputenc}
\usepackage{amsmath}
\usepackage{amssymb}
\usepackage{lipsum}
\usepackage{graphicx}
\usepackage{babel}

\usepackage{color}

\usepackage{ulem}

\newcommand{\stkout}[1]{\ifmmode\text{\sout{\ensuremath{#1}}}\else\sout{#1}\fi}

\begin{document}
\title{Superradiance from Nitrogen Vacancy Centers Coupled to An Ultranarrow Optical Cavity}

\author{Yi-Dan Qu}
\address{Henan Key Laboratory of Diamond Optoelectronic Materials and Devices, Key Laboratory of Material Physics Ministry of Education, School of Physics and Laboratory of Zhongyuan Light, Zhengzhou University, Daxue Road 75, Zhengzhou 450052, China}

\author{Yuan Zhang}
\email{yzhuaudipc@zzu.edu.cn}
\address{Henan Key Laboratory of Diamond Optoelectronic Materials and Devices, Key Laboratory of Material Physics Ministry of Education, School of Physics and Laboratory of Zhongyuan Light, Zhengzhou University, Daxue Road 75, Zhengzhou 450052, China}
\address{Institute of Quantum Materials and Physics, Henan Academy of Sciences, Mingli Road 266-38, Zhengzhou 450046, China}

\author{Peinan Ni}
\address{Henan Key Laboratory of Diamond Optoelectronic Materials and Devices, Key Laboratory of Material Physics Ministry of Education, School of Physics and Laboratory of Zhongyuan Light, Zhengzhou University, Daxue Road 75, Zhengzhou 450052, China}

\author{Chongxin Shan}
\email{cxshan@zzu.edu.cn}
\address{Henan Key Laboratory of Diamond Optoelectronic Materials and Devices, Key Laboratory of Material Physics Ministry of Education, School of Physics and Laboratory of Zhongyuan Light, Zhengzhou University, Daxue Road 75, Zhengzhou 450052, China}
\address{Institute of Quantum Materials and Physics, Henan Academy of Sciences, Mingli Road 266-38, Zhengzhou 450046, China}

\author{Hunger David}
\address{Physikalisches Institut, Karlsruhe Institute of Technology (KIT), Wolfgang-Gaede Str. 1, 76131
Karlsruhe, Germany}

\author{Klaus M\o lmer}
\email{klaus.molmer@nbi.ku.dk}
\address{Niels Bohr Institute, University of Copenhagen, Blegdamsvej 17, 2100 Copenhagen, Denmark}

\begin{abstract}
Nitrogen-vacancy (NV) centers in diamond have been successfully coupled to various optical structures to enhance their radiation by the Purcell effect. The participation of many NV centers in these studies may naturally lead to cooperative emission and superradiance, and our recent experimental study with a diamond membrane in a fiber-based ultra-narrow optical cavity demonstrated nonlinear radiation power and fast photon bunching which are signatures of such collective effects. In this theoretical article, we go beyond the simple model used in the previous study to address more phenomena, such as the appearance of bunching shoulders in the second-order correlation function, Rabi splitting in the steady-state spectrum, and  population dynamics on excited Dicke states, which for moderate pumping explains the observed collective effects. Overall, our results can guide further experiments with NV centers, and they are also relevant for other solid-state color centers, such as silicon-vacancy centers in diamond and silicon carbide, boron-vacancy centers and carbon-related centers in hexagonal boron-nitride.  
\end{abstract}
\maketitle

\section{Introduction}

Nitrogen-vacancy (NV) centers, is the prototypical solid-state spin system~\citep{WolfowiczG}, with rich electronic and spin levels, that can be explored by optical excitation of electronic transitions and  microwave excitation of spin transitions. Spin-preserving radiative decay and spin-sensitive inter-system crossings~\citep{DohertyMW} enable spin initialization with optical excitation, and spin readout with fluorescence~\citep{GruberA}, infrared absorption~\citep{AcostaVM} or photon-induced carriers~\citep{BourgeoisE}. The long coherence time of the spin transitions at room temperature, and the coupling with magnetic, electric and optical fields have promoted NV centers as quantum bits in quantum computing~\citep{PezzagnaS} and quantum information applications~\citep{WrachtrupJ},  in quantum sensing or precision  measurement applications~\citep{SegawaTF}, as well as gain materials in room-temperature maser~\citep{ArrooDM} and laser applications~\citep{SavvinA}.  

To enhance the coupling of the spin and optical transitions with the microwave and optical fields, NV centers have been integrated with various microwave structures, such as superconducting circuits~\citep{AmussR}, lumped elements~\citep{AngererA}, dielectric cavities~\citep{EisenachER}, and with optical structures, such as immersion lenses~\citep{RobledoL}, photonic crystals, optical and plasmonic cavities~\citep{BarthM}. By exploring the collective effects from billions of NV centers, many intriguing collective effects, such as superradiance~\citep{AngererA}, Rabi oscillations~\citep{PutzS}, Rabi splitting~\citep{AmussR,ZhangYPRL},  microwave cooling~\citep{NgW,ZhangYNPJ,FaheyDP}, and microwave amplification by stimulated emission~\citep{BreezeJD} have been demonstrated. However, the studies on the system integrated with optical structures have so-far focused on the enhanced radiation by the Purcell effect~\citep{RobledoL,BarthM}, while investigations of superradiance and strong coupling effects have been  impeded by NV center transitions being much broader than the ultranarrow resonances of the  optical structures, and hence restricting the systems to the weak coupling regime.

In our recently experimental study~\citep{PallmannM}, we showed that the frequency  fluctuations, i.e. the main cause of the zero-phonon line broadening, can be dramatically reduced by working at low temperature, where signatures of collective effects, i.e., a nonlinear dependence of the radiation power and the fast photon bunching, can be observed under the right conditions. To  theoretically account for the main physics in that study, we treated the NV centers as two-level systems subject to cavity-mediated collective decay, and showed that the NV centers are prepared to sub-radiant Dicke states under incoherent optical pumping, and are responsible for the observed collective effects. In the present article, we complement this study by developing more sophisticated models to account for the multi-level nature of the NV center, and we identify the singlet excited state to be responsible for the dependence of the photon bunching and anti-bunching on pumping and a shoulder in the bunching signal for longer time. Furthermore, we identify other interesting effects, such as scaling of the  radiation within different pumping regimes, and a  pumping-controlled optical Rabi splitting. Although intended here for NV centers, our results are also informative for other solid-state color center systems, such as silicon-vacancy centers in diamond~\citep{BeckerJN} and silicon carbide \citep{LohrmannA}, boron-vacancy centers~\citep{QianC} and carbon-related centers in hexagonal boron-nitride~\citep{DowranM}.

This article is organized as follows. In Sec.~\ref{sec:theory}, we present the theoretical models based on quantum master equations, and we discuss their solutions with the standard density matrix technique and the cumulant mean-field approach. In Sec. \ref{sec:results}, we present our numerical results, and we discuss separately the cases with few NV centers and with many NV centers. In Sec. IV, we summarize and conclude with an outlook of possible further studies.

\begin{figure}
\begin{centering}
\includegraphics[scale=0.31]{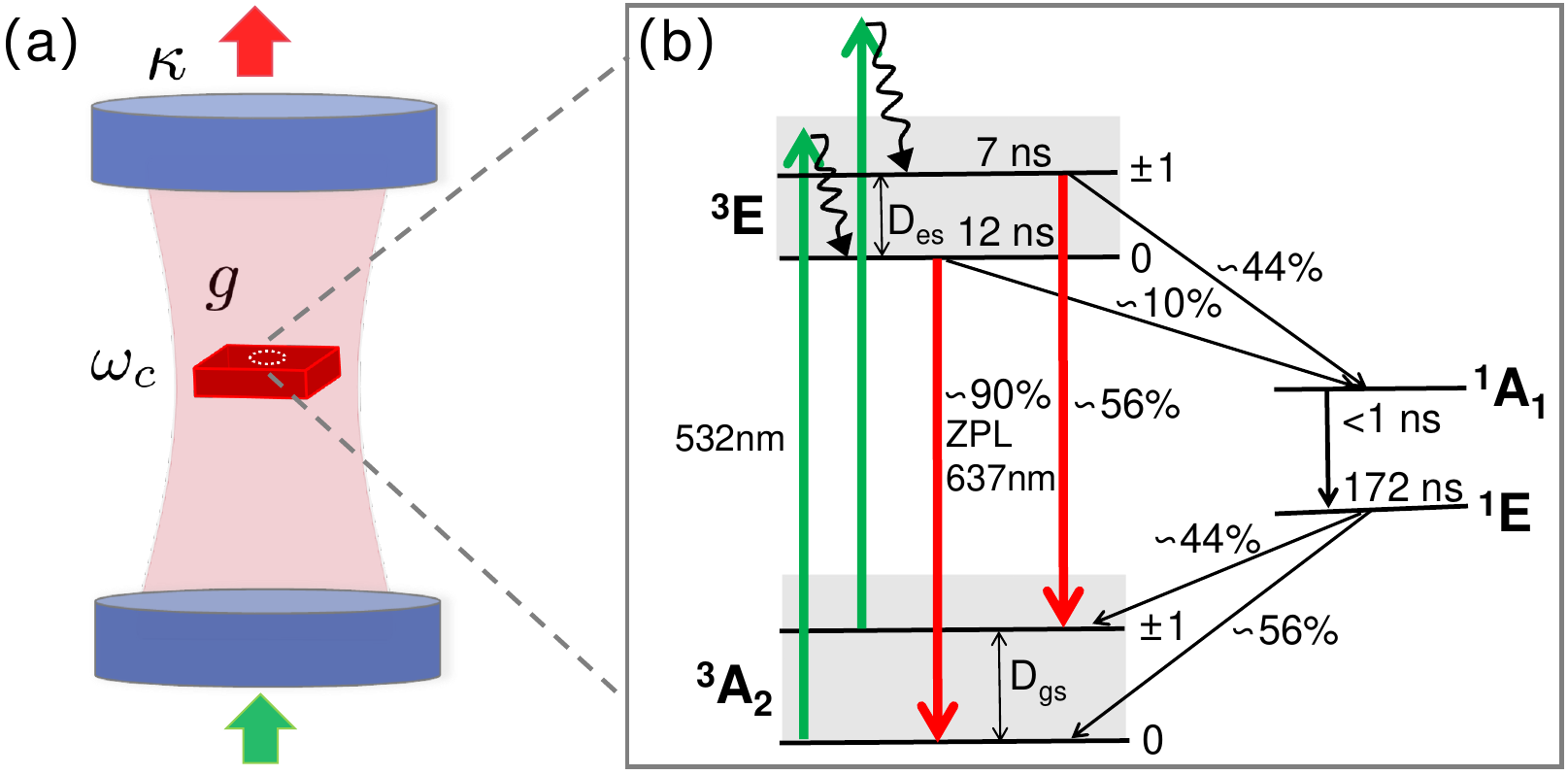}
\par\end{centering}
\caption{\label{fig:system} System schematics. Panel (a) shows the coupling between one fundamental mode of the ultranarrow optical cavity with a frequency $\omega_c$ and a photon damping rate $\kappa$ and many NV centers in a diamond membrane with a strength $g$, the incoherent pumping NV centers by a 532 nm laser through the cavity (green arrow), and the cavity-mediated florescence of the excited NV centers (red arrow). Panel (b) shows the simplified energy levels of the NV centers with two triplet levels $^3A_2$, $^3E$, which split into three spin levels $m_s = 0,\pm 1$ with  zero-field splitting $D_{gs}$,$D_{es}$, and two singlet excited levels $^1A_1$, $^1E$,  where the lifetimes and the branching ratios of decay channels are also indicated. The shaded areas show the phonon excitation, and the wave arrows indicate the fast phonon relaxation. }
\end{figure}

\section{Theoretical Models based on Quantum Master Equation \label{sec:theory}}

In our theory, we would like to model the dynamics of the diamond membrane-optical cavity system, as studied in the previous experiment~\citep{PallmannM} and schematically shown in Fig.~\ref{fig:system}(a). The diamond membrane locates within the waist of a fundamental cavity mode, and contains an ensemble of NV centers. Since the NV centers are generated by nitrogen-ion implementation, they  suffer from significant lattice strain, as collaborated by the broad zero-phonon line of hundreds GHz wide in the cavity-filtered fluorescence spectrum [see Fig. 2(b) of Ref.~\citep{PallmannM}]. Under such strong strain, it is expected that most of NV centers experience large level splitting among the excited electronic-spin levels, and thus they feature optical transitions with perpendicular polarizations \citep{MNB}. Since the optical cavity has a linewidth around GHz, most of NV centers couple with the cavity through one optical transition. Furthermore, the earlier study by Albrecht R. et. al., indicated that the Purcell-enhanced radiation can be achieved through the zero-phonon line transition of the NV centers, but not through the phonon-assisted transition due to the fast relaxation for the excited levels of the phonon mode~\citep{AlbrechtR}, which was also collaborated by our previous study. Although the phonon-assisted transitions might be important to reproduce quantitatively the experimental results, it is sufficient to describe phenomenologically the combined effect of the optical excitation and the phonon relaxation with an incoherent pumping rate. By following these considerations, we end up with a simplified model for the NV centers as shown in Fig.~\ref{fig:system}(b). Note that such a model is also usually adopted to investigate the optical detected magnetic resonance of the NV centers at room temperature~\citep{BusaiteL,DuarteH}.

\begin{figure}
\begin{centering}
\includegraphics[scale=0.35]{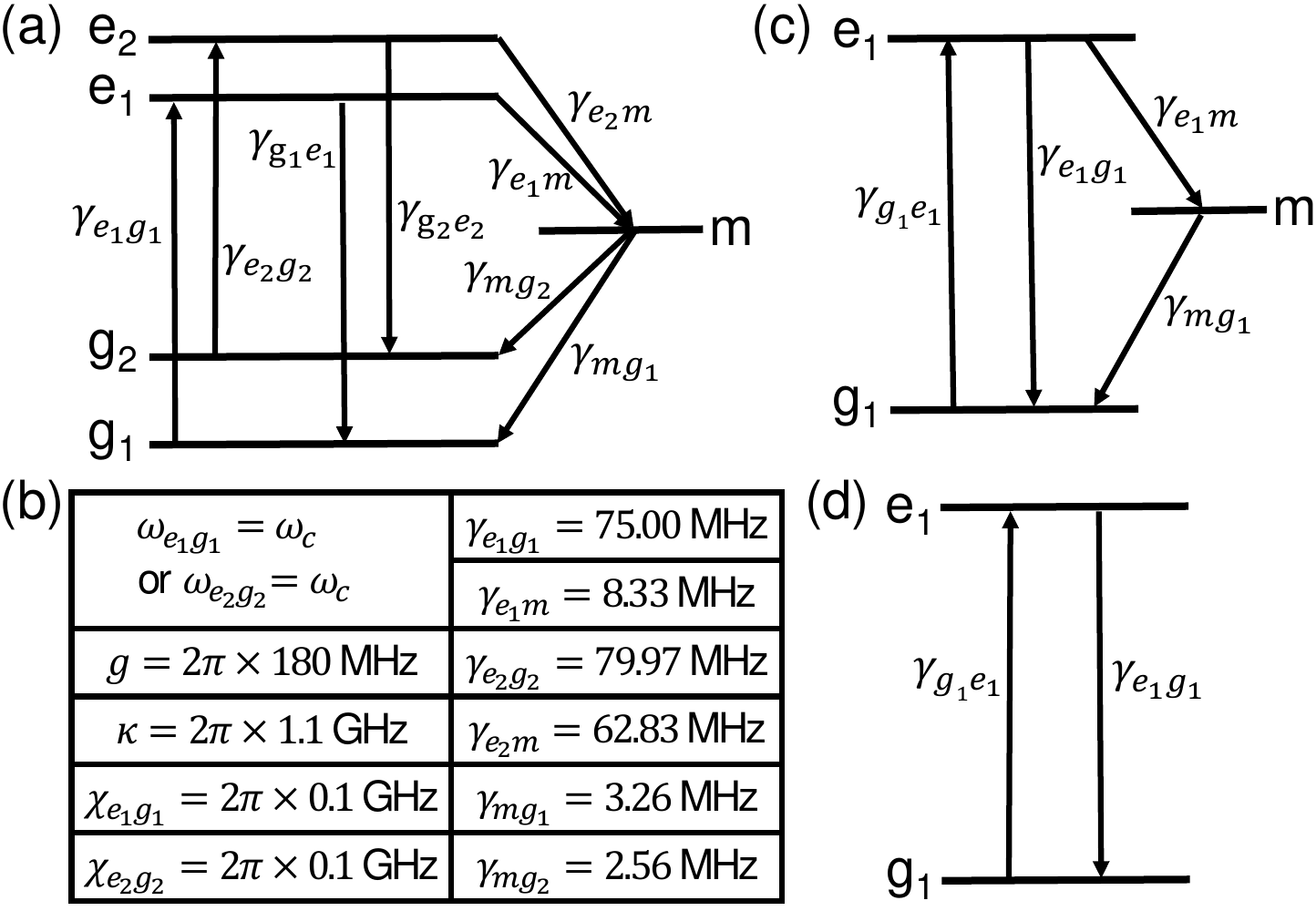}
\par\end{centering}
\caption{\label{fig:model} Effective models of NV centers. Panel (a) shows the effective five-level model, where the degenerated spin levels are grouped as single levels, and the two singlet levels are grouped as a single level, and the rates of different decays are marked. Panel (b) shows the parameters for the optical cavity and the NV centers. Panel (c) shows the simplified model, where the ground and excited levels with $m_s = \pm 1$  are ignored. Panel (d) shows the most simplified model, where the singlet excited levels are also ignored.}
\end{figure}

To account for the collective effects, we establish a theory within the framework of cavity quantum electrodynamics. We consider an optical cavity mode with a frequency $\omega_c$ and a photon damping rate $\kappa$, which couples to many NV centers with a strength $g$. The NV centers possess one triplet ground level $^3A_2$, one triplet excited level  $^3E$, and two singlet excited levels $^1A_1$, $^1E$. The two triplet levels split further into three spin levels, which are labeled with the projection number $m_s = -1,0,+1$ along the NV quantization axis, and the energy splitting between the level $m_s =0$ and the degenerated  levels $m_s = \pm 1$ are given by the so-called zero-field splitting $D_{gs}=2\pi\times2.87$ GHz and $D_{es}=2\pi\times1.42$ GHz. To simplify the the optical part of the energy diagram of the NV centers, we can group the degenerated spin levels $m_s=\pm1$ as one level due to their similar transition rates to other levels, and group the two singlet excited levels $^1A_1$, $^1E$ as one level due to the fast decay of the upper singlet excited level. As a result, we can model the NV centers as five-levels systems, and denote the spin levels $m_s=0,\pm1$ of the electronic level $^3A_2$ as the $g_1,g_2$ level, and the spin levels $m_s=0,\pm1$ of the electronic level $^3E$ as the $e_1, e_2$ level, and the singlet levels $^1A_1$, $^1E$ as the $m$ level [Fig. ~\ref{fig:model}(a)].  According to the life-time of different levels and the branching ratios as given in Fig.~\ref{fig:system}(b), we have also calculated the rates of different processes and present them in Fig.~\ref{fig:model}(b). Furthermore, we compare our model  with  two simplified ones, which treat the NV centers as three-level [Fig. \ref{fig:model}(c)] and two-level systems [Fig. \ref{fig:model}(d)], respectively, and demonstrate that they capture already the main collective effects as observed in the experiment~\citep{PallmannM}. 

In our theory, the system dynamics is described by the quantum master equation (QME) for the reduced density operator $\hat{\rho}$:
\begin{align}
 & \frac{\partial}{\partial t}\hat{\rho}=-\frac{i}{\hbar}[\hat{H}_{NV}+\hat{H}_{c}+\hat{H}_{NV-c},\hat{\rho}]-\kappa\mathcal{D}[\hat{a}]\hat{\rho} \nonumber \\ 
 & -\sum_{i=1}^{N}\{ \gamma_{e_{1}g_{1}}\mathcal{D}[\hat{\sigma}_{i}^{g_{1}e_{1}}]\hat{\rho}+\gamma_{g_{1}e_{1}}\mathcal{D}[\hat{\sigma}_{i}^{e_{1}g_{1}}]\hat{\rho} \nonumber \\
 & +(\chi_{e_{1}g_{1}}/2)\mathcal{D}[\hat{\sigma}_{i}^{e_{1}e_{1}}-\hat{\sigma}_{i}^{g_{1}g_{1}}]\hat{\rho}\} \nonumber\\
 & -\sum_{i=1}^{N}\{ \gamma_{e_{2}g_{2}}\mathcal{D}[\hat{\sigma}_{i}^{g_{2}e_{2}}]\hat{\rho}+\gamma_{g_{2}e_{2}}\mathcal{D}[\hat{\sigma}_{i}^{e_{2}g_{2}}]\hat{\rho} \nonumber \\
 &+(\chi_{e_{1}g_{1}}/2)\mathcal{D}[\hat{\sigma}_{i}^{e_{2}e_{2}}-\hat{\sigma}_{i}^{g_{2}g_{2}}]\hat{\rho}\} \nonumber\\
 & -\sum_{i=1}^{N}\{ \gamma_{e_{1}m}\mathcal{D}[\hat{\sigma}_{i}^{me_{1}}]\hat{\rho}+\gamma_{mg_{1}}\mathcal{D}[\hat{\sigma}_{i}^{g_{1}m}]\hat{\rho}\nonumber \\
 &+\gamma_{e_{2}m}\mathcal{D}[\hat{\sigma}_{i}^{me_{2}}]\hat{\rho}+\gamma_{mg_{2}}\mathcal{D}[\hat{\sigma}_{i}^{g_{2}m}]\hat{\rho}\}.
 \label{eq:qme}
 \end{align}
In this equation, we introduce the Hamiltonian $\hat{H}_{NV}=\hbar\sum_{i=1}^{N}[(\omega^{e_{1}g_{1}}/2)\left(\hat{\sigma}_{i}^{e_{1}e_{1}}-\hat{\sigma}_{i}^{g_{1}g_{1}}\right)+(\omega^{e_{2}g_{2}}/2)\left(\hat{\sigma}_{i}^{e_{2}e_{2}}-\hat{\sigma}_{i}^{g_{2}g_{2}}\right)]$ to describe the optical transitions of the NV centers with frequencies $\omega^{e_{1}g_{1}},\omega^{e_{2}g_{2}}$, which are coupled with the optical cavity. Here, the label $i=1,\ldots,N$ indicates the individual  of total $N$ NV centers, and $\hat{\sigma}_{i}^{e_{1}e_{1}},\hat{\sigma}_{i}^{g_{1}g_{1}},\hat{\sigma}_{i}^{e_{2}e_{2}},\hat{\sigma}_{i}^{g_{2}g_{2}}$ are the projection operators on the $e_{1},g_{1},e_{2},g_{2}$ levels.  The Hamiltonian $\hat{H}_{c}=\hbar\omega_{c}\hat{a}^{\dagger}\hat{a}$ describes the optical cavity with the frequency $\omega_{c}$, the photon creation and annihilation operators $\hat{a}^{\dagger},\hat{a}$.  The Hamiltonian $\hat{H}_{NV-c}=\hbar g \sum_{i=1}^{N} \left\{ \left(\hat{\sigma}_{i}^{e_{1}g_{1}}+\hat{\sigma}_{i}^{e_{2}g_{2}}\right)\hat{a}+\hat{a}^{\dagger}\left(\hat{\sigma}_{i}^{g_{1}e_{1}}+\hat{\sigma}_{i}^{g_{2}e_{2}}\right)\right\} $ describes the coherent energy exchange coupling between the NV optical transitions and the optical cavity with the strength $g$ and the transition operators $\hat{\sigma}_{i}^{e_{1}g_{1}},\hat{\sigma}_{i}^{e_{2}g_{2}},\hat{\sigma}_{i}^{g_{1}e_{1}}\hat{\sigma}_{i}^{g_{2}e_{2}}$.

The Lindblad term $\kappa\mathcal{D}[\hat{a}]\hat{\rho}$ on the first line of Eq. (\ref{eq:qme}) describes the cavity photon loss with a rate $\kappa$, and the superoperator $\mathcal{D}[\hat{o}]\hat{\rho}=1/2\left\{ \hat{o}^{\dagger}\hat{o},\hat{\rho}\right\} -\hat{o}\hat{\rho}\hat{o}^{\dagger}$ is defined for any operator $\hat{o}$. The second and third line of 
Eq. (\ref{eq:qme}) describe the spontaneous emission, the incoherent optical pumping, and the dephasing rates $\gamma_{e_{1}g_{1}},\gamma_{g_{1}e_{1}},\chi_{e_{1}g_{1}}$ for the $e_{1}\leftrightarrow g_{1}$  transition, and the fourth and fifth line describe the corresponding processes with rates $\gamma_{e_{2}g_{2}},\gamma_{g_{2}e_{2}},\chi_{e_{2}g_{2}}$ for the $e_{2}\leftrightarrow g_{2}$ transition. Since the optical pumping of NV centers under the laser illumination involves also the phonon excitation and the fast phonon relaxation, we treat this process phenomenologically with single rate. 
The last line of Eq. (\ref{eq:qme}) describes the inter-system crossings from the excited triplet levels to the singlet levels, and from the singlet levels to the ground triplet levels for the spin $m_{s}=0$ ($m_{s}=\pm1$) with the rates $\gamma_{e_{1}m},\gamma_{mg_{1}} (\gamma_{e_{2}m},\gamma_{mg_{2}}$).

In the above theory, we have assumed that the rates and the coupling are same for all the NV centers, which introduces symmetry to the system and reduces dramatically the independent elements (see below for more details). In addition, we have also ignored the relatively slow spin dephasing and the spin-lattice relaxation for the sake of simplicity. In the following, we present the density matrix method and the cumulant mean-field approach to solve the QME~\eqref{eq:qme}. 

\subsection{Solutions based on Density Matrix Method}

In the standard density matrix method, we introduce the product states $\left|\alpha\right\rangle =\left|a\right\rangle \left|n\right\rangle $ with the electron-spin states $\left|a\right\rangle =\left\{ \left|g_{1}\right\rangle ,\left|g_{2}\right\rangle ,\left|e_{1}\right\rangle ,\left|e_{2}\right\rangle ,\left|m\right\rangle \right\} $ and the photon number states $\left|n\right\rangle $ ($n=0,1,2,\ldots$ for vacuum, single and multiple photon states), and define the density matrix with elements $\rho_{\alpha\beta}=\left\langle \alpha\right|\hat{\rho}\left|\beta\right\rangle $ in the Hilbert space spanned by these product states, and finally derive equation for the matrix elements from the QME~\eqref{eq:qme}. In practice, we carry out the above procedure with the QuTiP package~\citep{JohanssonJR}, and present the corresponding Python codes in Fig. \ref{fig:PythonCodes}. Unfortunately, the density matrix technique can only be applied to the system with few NV centers, due to the exponentially increased Hilbert space. To consider the system with more NV centers, we can further simplify the models and consider the NV centers as the three-level and two-level systems, as examined below. 

In the model treating the NV centers as two-level systems, it is also possible to explore the permutation symmetry of identical particles. To this end, we simplify and rewrite the QME~\eqref{eq:qme} as follows
\begin{align}
 & \frac{\partial}{\partial t}\hat{\rho}=-\frac{i}{\hbar}[\hat{H}_{NV}+\hat{H}_{c}+\hat{H}_{NV-c},\hat{\rho}]-\kappa\mathcal{D}[\hat{a}]\hat{\rho} \nonumber \\ 
 & -\sum_{i=1}^{N}\{\gamma_{e_{1}g_{1}}\mathcal{D}[\hat{\sigma}_{i}^-]\hat{\rho}+\gamma_{g_{1}e_{1}}\mathcal{D}[\hat{\sigma}_{i}^+]\hat{\rho} +(\chi_{e_{1}g_{1}}/2)\mathcal{D}[\hat{\sigma}_i^z]\hat{\rho}\}, \label{eq:qme-tls}
\end{align}
where the Hamiltonians become $\hat{H}_{NV}=(\omega^{e_1 g_1}/2)\sum_{i=1}^N
\hat{\sigma}_{i}^z$ and $\hat{H}_{NV-c}= \hbar g \sum_{i=1}^N
(\hat{\sigma}_{i}^+\hat{a} + \hat{a}^\dagger\hat{\sigma}_{i}^-)$. Here, we have assumed that the optical cavity couples with the NV centers through the optical transition  with $m_s=0$, and introduced the operators $\hat{\sigma}_{i}^- =   \hat{\sigma}_{i}^{g_{1}e_{1}}$, $\hat{\sigma}_{i}^+ =   \hat{\sigma}_{i}^{e_{1}g_{1}}$ and $\hat{\sigma}_i^z = \hat{\sigma}_{i}^{e_{1}e_{1}}-\hat{\sigma}_{i}^{g_{1}g_{1}}$.  To proceed, we introduce the collective spin operators $\hat{J}_x = (1/2)\sum_{i=1}^N (\hat{\sigma}_{i}^- + \hat{\sigma}_{i}^+)$, $\hat{J}_y = (i/2)\sum_{i=1}^N (\hat{\sigma}_{i}^- - \hat{\sigma}_{i}^+)$ and $\hat{J}_z = (1/2)\sum_{i=1}^N \hat{\sigma}_{i}^z$, and then define the Dicke states as the eigen states of two collective operators: $(\hat{J}_x^2+\hat{J}_y^2 + \hat{J}_z^2)\left| J, M\right \rangle = J(J+1) \left| J, M\right \rangle$ and $\hat{J}_z \left| J, M\right \rangle = M \left| J, M\right \rangle$~\citep{RHDicke}. Here, the integers or half-integers $J=-N/2,...,N/2 $ characterize the coupling strength of the NV centers to the cavity mode, and the numbers $M=-J,...,J$ characterize the excitation of the NV center ensemble. Usually, the Dicke states for given $J$ form a ladder with equal spacing, and the states for different $J$ are visualized as shifted ladders, forming a triangle space [Fig.~\ref{fig:Dicke}(a-c)].  Furthermore, by using Clebsch-Gordan expansion to represent the Dicke states with given $J$ as the product of the Dicke states with $J-1$ and the states of single two-level system, it is also possible to represent the individual dissipations given by the second line of Eq.~\eqref{eq:qme-tls} as the quantum jumps between different Dicke states~\citep{BaragiolaBQ,ZhangYNJP}. As a result, it is possible to define the density matrix within the Dicke states, and translate Eq.~\eqref{eq:qme-tls} as a matrix equation. In practice, N. Shammah et. al. have developed a package based on Python language~\citep{ShammahN} to implement such an equation for more general problem. In Fig.~\ref{fig:PythonCodesDicke}, we present the corresponding codes for our problem. 

\subsection{Solutions based on Cumulant Mean-field Approach}

To consider the system with more NV centers,  we can also explore the cumulant mean-field approach. In this approach, we derive the equations $\partial_t \left\langle \hat{o}\right\rangle =\mathrm{tr}\left\{ \hat{o}\partial_t \hat{\rho}\right\} $ for the mean fields $\left\langle \hat{o}\right\rangle =\mathrm{tr}\left\{ \hat{o}\hat{\rho}\right\}$ of  operators $\hat{o}$ from the QME  (\ref{eq:qme}), and find that they depend on the mean fields of double operators $\left\langle \hat{o}\hat{p}\right\rangle $, which further depend on the mean fields of three operators $\left\langle \hat{o}\hat{p}\hat{q}\right\rangle $ and lead to a hierarchy of equations. To truncate the hierarchy, we can utilize the cumulant expansion approximations, e.g. $\left\langle \hat{o}\hat{p}\hat{q}\right\rangle \approx\left\langle \hat{o}\right\rangle \left\langle \hat{p}\hat{q}\right\rangle +\left\langle \hat{p}\right\rangle \left\langle \hat{o}\hat{q}\right\rangle +\left\langle \hat{q}\right\rangle \left\langle \hat{o}\hat{p}\right\rangle -2\left\langle \hat{o}\right\rangle \left\langle \hat{p}\right\rangle \left\langle \hat{q}\right\rangle $, to obtain a closed set of equations. In addition, we can also assume that all the NV centers are identical, and reduce the number of equations dramatically by removing the resulted degeneracy of the mean-fields. In practice, we employ the QuantumCumulants.jl package~\citep{PlankensteinerD} to implement the mean-field treatment (see the Appendix ~\ref{sec:mf}).

To better illustrate the collective dynamics, we follow our previous studies~\citep{ZhangYPRL,ZhangYNPJ} to convert the mean values to  the average of Dicke states quantum numbers $\overline{J},\overline{M}$. If we indicate the two levels coupled with the optical cavity with $1,2$, these numbers can be computed as $\overline{M}=(N/2)(\left\langle \hat{\sigma}_{1}^{22}\right\rangle-\left\langle\hat{\sigma}_{1}^{11}\right\rangle )$, $\overline{J}=[1+\sqrt{1+4(J_{x}^{2}+J_{y}^{2}+J_{z}^{2})}]/2-1$, with $J_{x(y)}^{2}=(N/4)[\pm1+(N-1)(\left\langle\hat{\sigma}_{1}^{12}\hat{\sigma}_{2}^{21}\right\rangle\pm\left\langle\hat{\sigma}_{1}^{21}\hat{\sigma}_{2}^{21}\right\rangle\pm\left\langle\hat{\sigma}_{1}^{12}\hat{\sigma}_{2}^{12}\right\rangle+\left\langle\hat{\sigma}_{1}^{21}\hat{\sigma}_{2}^{12}\right\rangle )]$, $J_{z}^{2}=(N/4)[(N-1)(\left\langle\hat{\sigma}_{1}^{22}\hat{\sigma}_{2}^{22}\right\rangle-2\left\langle \hat{\sigma}_{1}^{11}\hat{\sigma}_{2}^{22}\right\rangle+\left\langle\hat{\sigma}_{1}^{11}\hat{\sigma}_{2}^{11}\right\rangle)+\left\langle\hat{\sigma}_{1}^{11}\hat{\sigma}_{2}^{11}\right\rangle+\left\langle\hat{\sigma}_{1}^{22}\hat{\sigma}_{2}^{22}\right\rangle]$. In these expressions, $\langle \hat{\sigma}_{1}^{ii} \rangle$ are the population on the two levels, $\langle \hat{\sigma}_{1}^{ij}\hat{\sigma}_{2}^{mn} \rangle$ are the correlation between the two representative NV centers (with $i,j,m,n$ representing $1$ and $2$).

After solving the QME~\eqref{eq:qme}, we can calculate several observables of interest, e.g., the population of  various levels $\left \langle \hat{\sigma}_{i}^{ll} \right \rangle (t)$ (with $l=g_1,g_2,e_1,e_2,m$), and the radiation rate $I_{rad}=\kappa 
 \left \langle \hat{a}^+\hat{a} \right \rangle (t)$ with the intra-cavity mean photon number $\left \langle \hat{a}^+\hat{a} \right \rangle$. In addition, we can also
 apply the quantum regression theorem~\citep{CGardiner2004} to calculate  the radiation spectrum $S(\omega)= 2\kappa{\rm Re}\int d \tau e^{-i\omega\tau} \left \langle \hat{a}(\tau+t_{ss})\hat{a}(t_{ss}) \right \rangle$, and  the second-order auto-correlation function $g^{\left(2\right)}\left(\tau\right)=\left\langle \hat{a}^{\dagger}(t_{ss})\hat{a}^{\dagger}(t_{ss}+\tau)\hat{a}(t_{ss}+\tau)\hat{a}(t_{ss})\right\rangle /[I_{rad}(t_{ss})]^{2}$, where the time argument $\tau$ refers to the time difference from the steady-state at $t_{ss}$ (see the Appendix \ref{sec:speccorr} for more details).

\section{Results \label{sec:results}}

\begin{figure}
\begin{centering}
\includegraphics[scale=0.17]{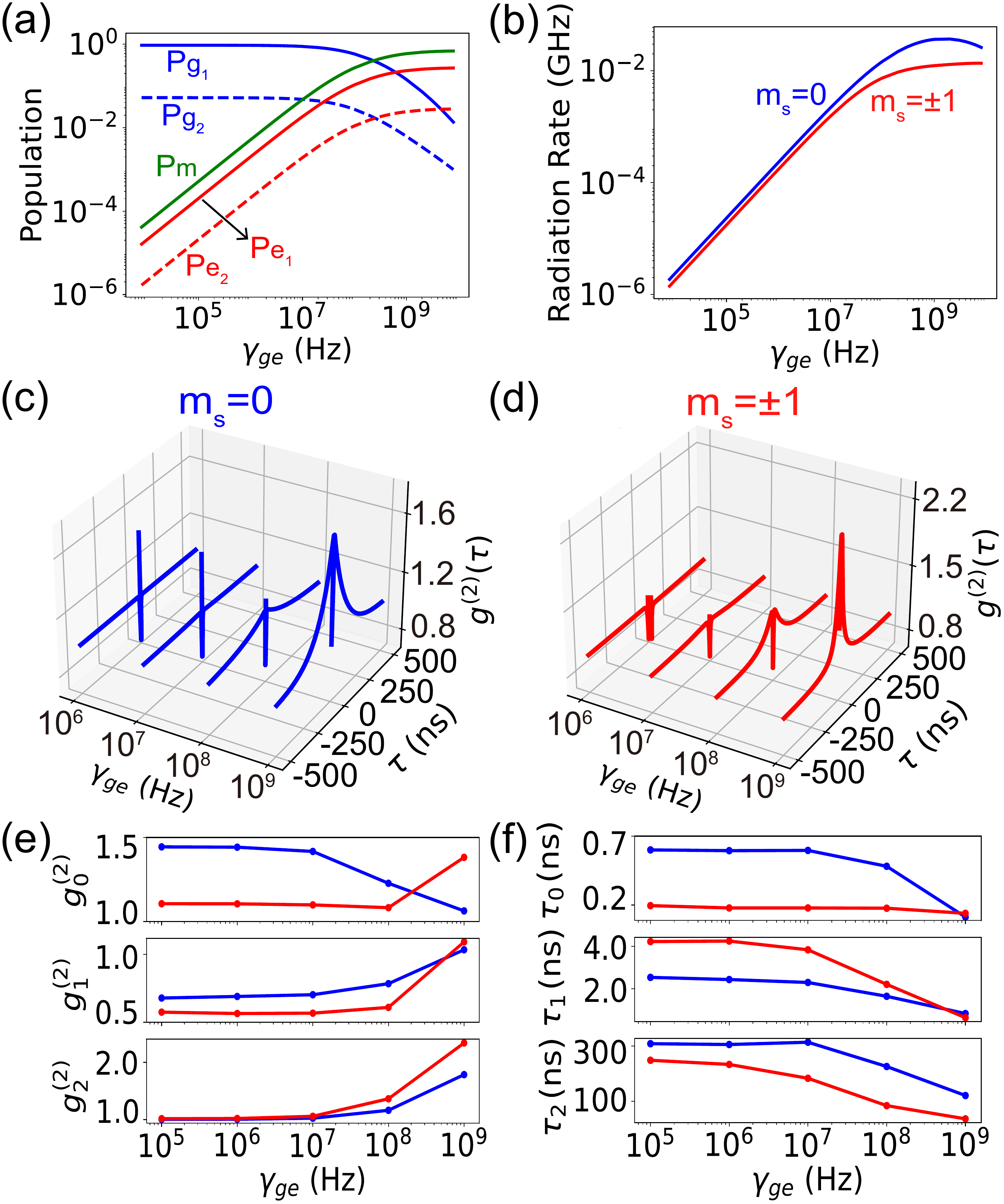}
\par\end{centering}
\caption{Dynamics of two five-level NV centers coupled to an optical cavity. Panel (a) shows the population of various levels as a function of the optical pumping rate $\gamma_{ge}=\gamma_{g_1 e_1}=\gamma_{g_2 e_2}$. Panel (b) shows the radiation rate for the situations, where the optical transitions with $m_s=0$ and $\pm1$  are resonant with the optical cavity  (blue and red line), respectively.  Panel (c) and (d) show the $g^{(2)} (\tau)$ function for increasing $\gamma_{ge}$ in the two cases. Panel (e) and (f) show the characteristic values $g_0^{(2)},g_1^{(2)},g_2^{(2)}$, and time $\tau_0,\tau_1,\tau_2$ of $g^{(2)} (\tau)$ as function of $\gamma_{ge}$, respectively. The former three parameters indicate the maximum and minimum of the features near zero delay time, and the maximum of the bunching shoulders. The latter three parameters denote the decay and raising time of the features near zero delay time, and the decay time of the bunching shoulders. Here, the results are computed with the density matrix technique. \label{fig:few}}
\end{figure}

In this section, we present  our numerical results on the systems with two, few and many NV centers, which are obtained with the density matrix method and the mean-field approach, respectively,  to reveal the emergence of the collective effects. In the end, we show the results based on the density matrix method in the Dicke states space to reveal more insights into the involved physics.  

\subsection{Systems with Two NV Centers}

We examine firstly the system with two NV centers, which are considered as five-level systems and treated with the density matrix approach (Fig. \ref{fig:few}). We examine the population of various levels as a function of the optical pumping rate $\gamma_{ge}=\gamma_{g_i e_i}$ ($i=1,2$) in Fig. \ref{fig:few}(a). We see that the population of the excited levels increases initially linearly, and then sub-linearly, and finally becomes saturated with increasing  $\gamma_{ge}$. In contrast, the population of the ground levels remain unchanged for weak pumping, and reduces dramatically for strong pumping. In addition, the populations of the lower ground and excited levels  are always larger than that of the upper ones, and the population is mostly trapped at the meta-stable level for the strong pumping. The population evolution of the two excited levels lead to the similar dynamics of the radiation rate [Fig. \ref{fig:few}(b)]. Since the population of the lower excited level is about ten times larger than that of the upper excited level, the radiation rate is much larger if the optical cavity is resonant to the NV optical transition with $m_s=0$. This indicates the dominating emission from this transition, and also motivates us to establish the simplified model to treat the NV centers as three-level systems [Fig.~\ref{fig:model}(c)], as elaborated further below.  

We now turn to the $g^{(2)} (\tau)$ function as a function of the optical pumping rate $\gamma_{ge}$ [Fig. \ref{fig:few}(c) and (d)]. In general, the $g^{(2)} (\tau)$ function shows one sharp bunching peak at zero delay time, and two anti-bunching dips at slightly later time, as well as two bunching shoulders for longer time. The sharp bunching peak does not occur in the typical $g^{(2)} (\tau)$ function of NV centers at room temperature, and is attributed to the cavity-mediated collective emission of multiple emitters~\citep{BradacC,LukinDM}. For increasing $\gamma_{ge}$,  the sharp peak decreases, while the dips and the shoulders increase when the cavity is resonant to the $m_s=0$ transition [Fig. \ref{fig:few}(c)]. However, all the quantities increase with increasing $\gamma_{ge}$ in the case of the $m_s=\pm 1$ transition [Fig. \ref{fig:few}(d)]. To quantify these changes, we estimate the maximum of the bunching peaks $g_0^{(2)}$, the minimum of the anti-bunching dips $g_1^{(2)}$, and the maximum of the bunching shoulders $g_2^{(2)}$ [Fig. \ref{fig:few}(e)]. In the case of the $m_s=0$ transition, $g_0^{(2)},g_1^{(2)},g_2^{(2)}$ remain as the constant values $1.4,0.68,1.0$ for $\gamma_{ge}<10^7$ Hz, and $g_0^{(2)}$ starts decreasing while the other two start increasing for $\gamma_{ge}>10^7$ Hz (blue lines). In the case of the $m_s=\pm1$ transition, the three quantities have the constant values $1.1,0.55,1.0$ for $\gamma_{ge}<10^7$ Hz, and they all start increasing for $\gamma_{ge}>10^8$ Hz (red lines). Here, the pumping rates, leading to the changes of $g^{(2)} (\tau)$, coincide with those leading to the changes of the ground levels population, as shown in Fig. \ref{fig:few}(a).

Furthermore, we determine the decay time $\tau_0$ of the sharp bunching peaks, the raising time $\tau_1$ and the decay time $\tau_2$ of the bunching shoulders [Fig. \ref{fig:few}(f)]. We find that $\tau_0$ is better estimated by fitting the calculated $g^{(2)} (\tau)$ curves with a Gaussian function instead of an exponential function. If the cavity is resonant to the NV optical transition with $m_s=0$, $\tau_0,\tau_1,\tau_2$ are within sub-, few- and hundreds of nanoseconds for $\gamma_{ge}<10^7$ Hz, and start decreasing for the larger pumping rate $\gamma_{ge}>10^7$ Hz (blue lines). Since $\tau_0 \approx 0.6$ ns for weak pumping is close to the inverse of the collective decay rate $1/(N\Gamma_c) \approx 0.74$ ns (for $N=2$), the sharp bunching peak can be attributed to the cavity-mediated collective radiation. Here, $\Gamma_c=4g^2/\kappa\approx 0.74$ GHz is the cavity-mediated decay rate of single NV center. Since $\tau_1\approx 2.5$ ns for the weak pumping is the same order as the lifetime of the upper excited level, the bunching feature can be attributed to the decay of this level. However, in the presence of the Purcell-enhanced decay, the actual value is about several times smaller than the lifetime of the upper excited level. Since $\tau_2\approx 300$ ns is the same order as the lifetime $172$ ns of the meta stable level, we can attribute the bunching shoulder to the decay of such a level. For the cavity resonant to the NV optical transition with $m_s = ±1$, the time $\tau_0,\tau_1,\tau_2$ behave similarly, except that $\tau_0$ remains unchanged for all the pumping rate. We note that the results achieved here are slightly different from what expected from the existing model treating NV centers as three-level systems~\citep{AlbrechtR,KurtsieferC}. Such a difference might be attributed to the influence of the cavity, and will be clarified in future. 


\begin{figure}
\begin{centering}
\includegraphics[scale=0.32]{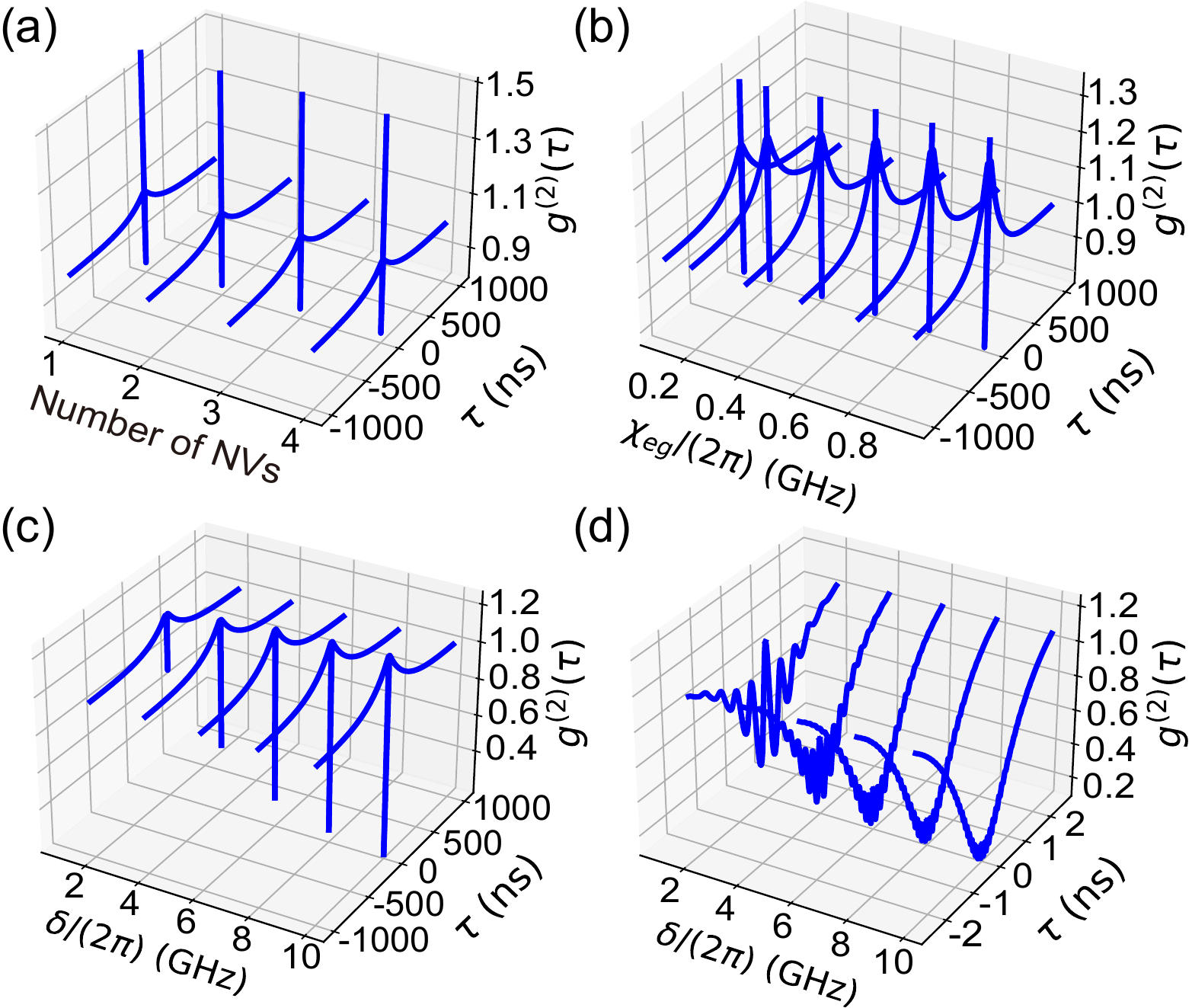}
\par\end{centering}
\caption{ \label{fig:3modelg2} $g^{(2)} (\tau)$ function calculated with the three-level model as function of the number of NV centers (a), the dephasing rate (b), and the frequency detuning to the cavity (c), respectively. Panel (d) shows the zoom-in of the panel (c) in the short time range. Here, we assume the optical pumping rate as $\gamma_{ge} =75$ MHz, and the results are computed with the density matrix technique. }
\end{figure}

\subsection{Systems with Few NV Centers}

The above results suggest that the radiation should be dominated by the optical transition with $m_s=0$, which motivates us to further simplify the NV centers as the three-level [Fig. \ref{fig:model}(c)] and the two-level systems [Fig. \ref{fig:model}(d)]. We have verified in Fig. \ref{fig:23models} (a,b) that the three-level model reproduces all the trend of the radiation rate, the level populations, and the $g^{(2)} (\tau)$ function. Motivated by these results, we have relied this model to further investigate the influence of other parameters on the $g^{(2)} (\tau)$ function (Fig. \ref{fig:3modelg2}). We find that as the number of NV centers increases from two to four, the maximum and minimum of the $g^{(2)} (\tau)$ function remain almost unchanged, and the bunching shoulders decrease slightly [Fig. \ref{fig:3modelg2}(a)]. As the dephasing rate of the NV centers increases, the bunching shoulders of the $g^{(2)} (\tau)$ function increase while other features remain unaffected [Fig. \ref{fig:3modelg2}(b)]. To study the influence of the frequency detuning to the cavity, we assume that one NV center is resonant to the cavity, and the other two are detuned gradually from the cavity with the same amount. We see that as the frequency detuning  increases over the cavity damping rate, the sharp bunching peak reduces below one, and becomes eventually smaller than $0.5$  [Fig. \ref{fig:3modelg2}(c)]. This indicates that there is effectively only single NV center coupled with the optical cavity  for large frequency detunning. In addition, for large frequency detuning, the $g^{(2)} (\tau)$ function shows also drastic oscillations which might be attributed to the cavity-mediated coupling between the NV centers [Fig. \ref{fig:3modelg2}(d)]. 

We have also verified in Fig. \ref{fig:23models} (c,d) that the two-levels model reproduces most of the features except for the bunching shoulders in the $g^{(2)} (\tau)$ function, which is known to originate from the singlet excited levels. Despite of this shortcoming, the two-level model can be employed to simulate the system with many NV centers by exploring the density matrix in the Dicke states space~\citep{ShammahN}, and to reveal the physics leading to the non-linear radiation and the sharp photon bunching effects, as observed in Ref.~\citep{PallmannM} and demonstrated in Sec.~\ref{sec:Dicke}.

\begin{figure}
\begin{centering}
\includegraphics[scale=0.25]{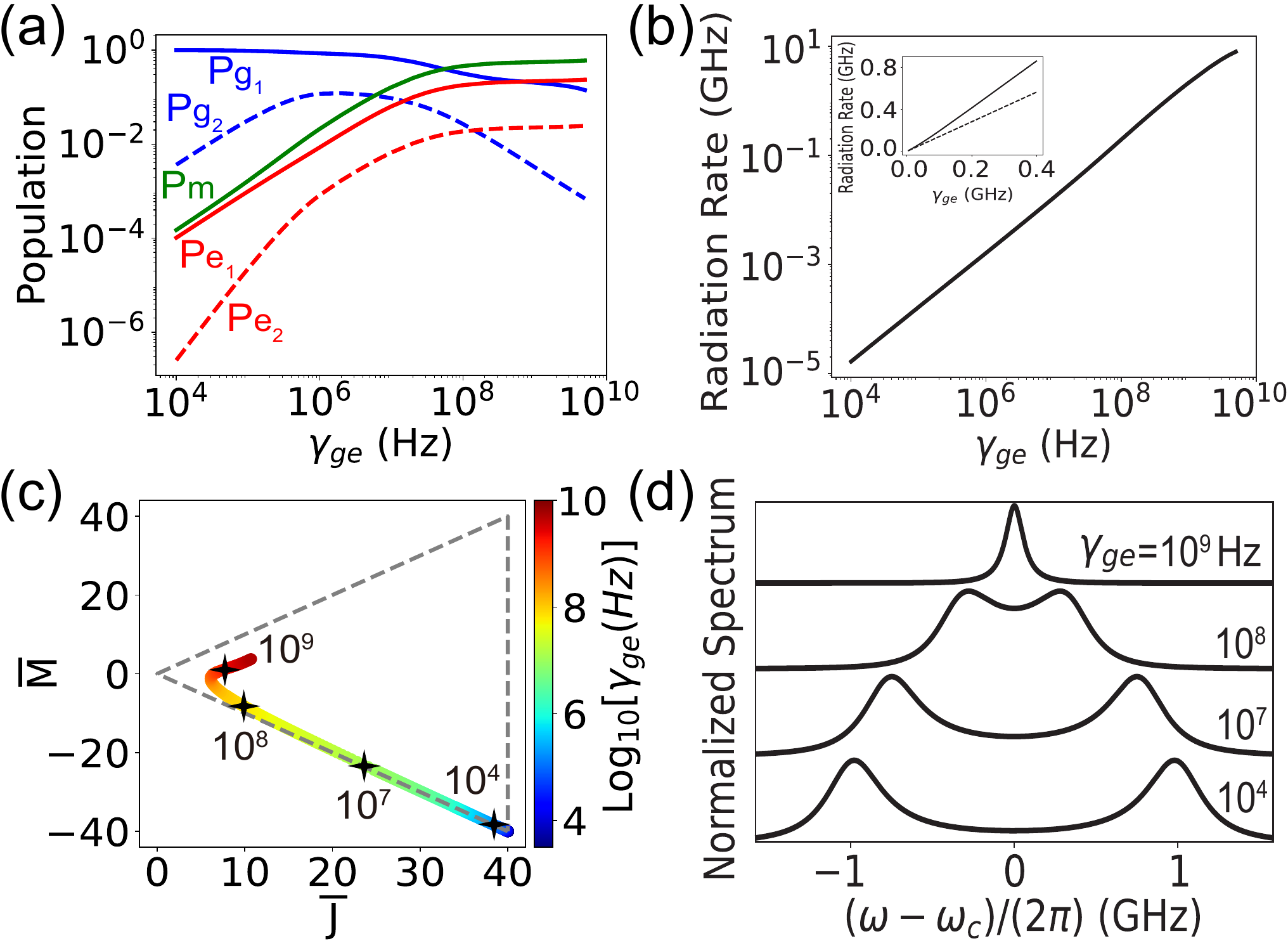}
\par\end{centering}
 \caption{Steady-state evolution of the system with $80$ NV centers as function of the optical pumping rate $\gamma_{ge}$.  Panel (a-c) show the population of various NV levels (a), the radiation rate (b), and the average of the Dicke states quantum numbers $\overline{J},\overline{M}$ (c). In the panel (b), the inset shows the zoom-in at moderate pumping, which shows a super-linear scaling. In the panel (c), the gray dashed lines indicate the boundaries of the Dicke states space. Panel (d) shows the spectrum for four representative pumping rates, as marked in the panel (c). Here, the results are computed with mean-field approach. Note that $g^{(2)} (\tau)$ function is excluded here since the mean-field approach can not be faithfully applied to compute this quantity so far. \label{fig:moreNVs}}
\end{figure}

\subsection{\label{sec:dickedm}Systems with Many NV Centers}

The above results indicate that the collective effect appears only in the $g^{(2)} (\tau)$ function for the system with few NV centers. To observe the collective effects directly from the radiation intensity or spectrum, we have to consider the system with more NV centers to enhance the coupling. In  Fig. \ref{fig:moreNVs}, we examine the steady-state evolution of the system with $80$ NV centers as a function of the optical pumping rate. To obtain these results, we treat the NV centers as five-level systems and solve the QME (\ref{eq:qme}) with the cumulant mean-field approach. We find that the level populations of  the NV centers  [Fig. \ref{fig:moreNVs} (a)] behave similarly as those for the system with few NV centers [Fig. \ref{fig:few} (a)], except that the population of the lower excited level $e_1$ approaches and overcomes that of the lower ground level $g_1$ for strong pumping. Accompanying with this population change, the radiation rate shows the transition from linear to super-linear and finally to sub-linear scaling for the weak, intermediate and strong pumping rate [Fig. \ref{fig:moreNVs} (b)], where the super-linear scaling is more clearly illustrated in the zoom-in plot (inset).

\begin{figure}
\begin{centering}
\includegraphics[scale=0.35]{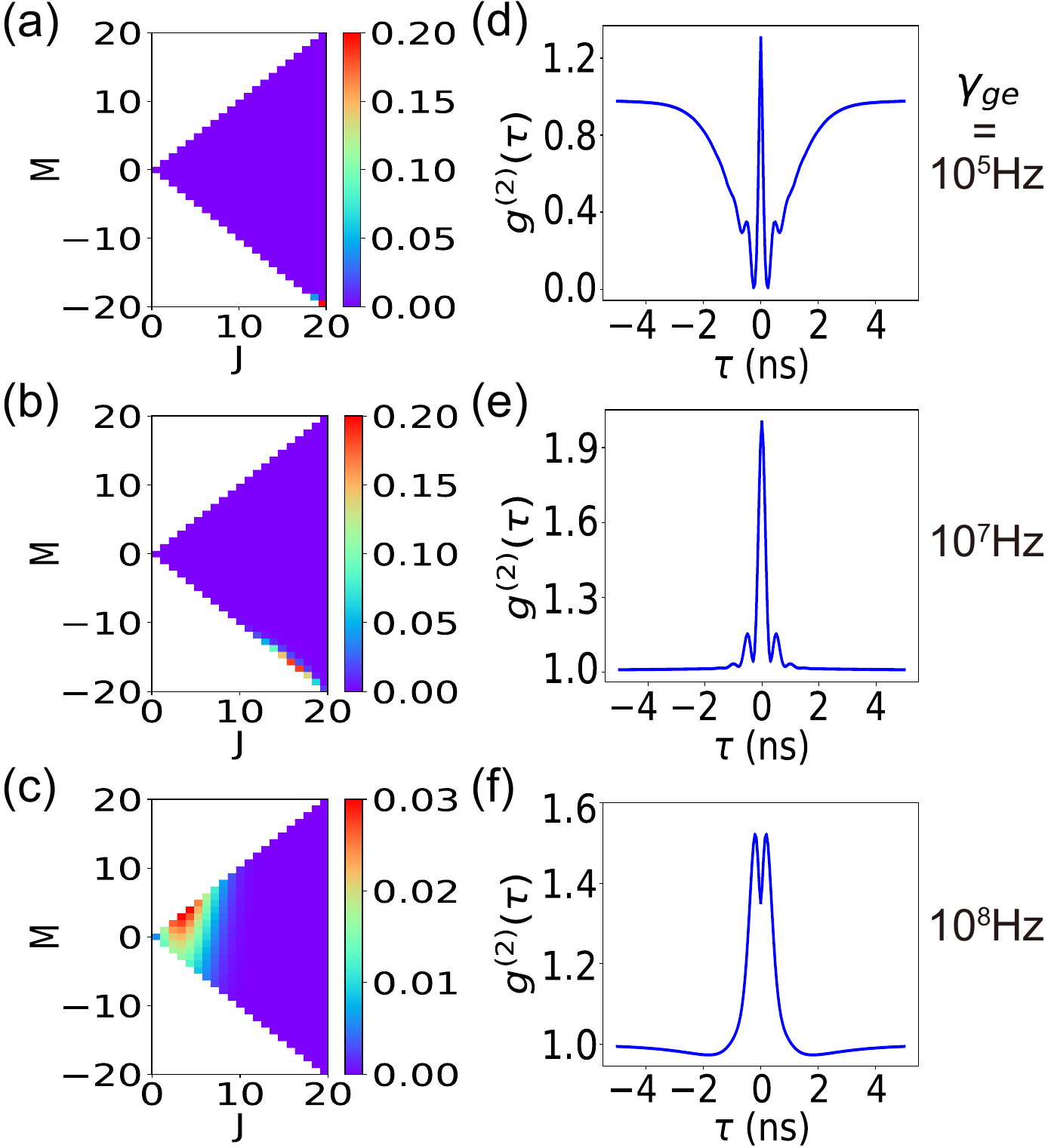}
\par\end{centering}
 \caption{Evolution of the Dicke states population (a-c) and the $g^{(2)} (\tau)$-function (d-f) of system with $40$ NV centers for increasing pumping rates $\gamma_{ge}=10^5,10^7,10^8$ Hz (from upper to lower panels). The evolution of radiation rate with $\gamma_{ge}$ is shown in Fig. A3(a), and the superlinear radiation occurs for $\gamma_{ge}>10^7$ Hz. The results are computed with the density matrix in the Dicke states space. \label{fig:Dicke}}
\end{figure}

To understand the super-linear scaling, we have computed the mean value of the Dicke states quantum numbers for different pumping rates [Fig. \ref{fig:moreNVs}(c)]. We see that the ensemble occupies the states near to lower-right corner for weak pumping, climbs along the lower boundary for increased pumping, departs from the boundary for further increased pumping, approaches the states with $\overline{M}=0$ and $\overline{J}\approx 6$, and climbs eventually along a line parallel to the upper boundary for much larger pumping. Here, the NV ensemble does not follow the upper boundary for large pumping because part of NV centers are pumped to the meta-stable level, and do not contribute to the definition of the Dicke states. We now utilize the occupation of the Dicke states to explain the behavior of the radiation. With increasing incoherent pumping, the NV ensemble also starts occupying the excited Dicke states with $\overline{M}>-\overline{J}$, and they can emit multiple photons through the cavity-mediated emission, leading to the superradiance. Indeed, the superlinear scaling of the radiation occurs when the NV ensemble starts departing from the lower boundary. This explanation becomes clearer in Fig.~\ref{fig:Dicke}, where we treat the NV centers as two-level systems and consider their dynamics with the population of the Dicke states.

Besides of the nonlinear radiation, we further examine the steady-state spectrum [Fig. \ref{fig:moreNVs}(d)] for increasing pumping rate. We observe two peaks at the frequencies about $2\pi$ GHz relative to the optical cavity for the weak pumping, and the approach of the two peaks for moderate pumping, and finally the merging of two peaks for the largest pumping rate. To understand the origin of these peaks, we examine again the Dicke states occupied by the NV ensemble [Fig. \ref{fig:moreNVs}(c)]. For the weak and moderate pumping, the NV ensemble occupies the Dicke states near to the lower boundary, and is not excited too much for given $J$. In this case, it is valid to apply the Holstein-Primakoff approximation to approximate these states as occupation number states of a quantized harmonic oscillator~\citep{HolsteinT}. As a result, the system Hamiltonian $\hat{H}=\hat{H}_{NV}+\hat{H}_c+\hat{H}_{NV-c}$ can be approximated as $\hbar \omega_{e_1 g_1}\hat{b}^\dagger \hat{b} + \hbar \omega_c \hat{a}^\dagger \hat{a} + \hbar g \sqrt{2J}(\hat{b}^\dagger \hat{a} + \hat{a}^\dagger \hat{b})$ with the creation and annihilation operator $\hat{b}^\dagger,\hat{b}$ of the oscillator. Here, we assume that the cavity is resonant with the NV optical transition with $m_s = 0$, and have ignored the general energy shift for the simplicity. By diagonalizing such a Hamiltonian as $\hat{H}\approx \sum_{\alpha=\pm} \hbar \omega_{\alpha} \hat{c}^\dagger_\alpha \hat{c}_\alpha$  with $\omega_\pm = \omega_c \pm \sqrt{J}g_1$ and $\hat{c}_{\alpha=\pm} =1/\sqrt{2}(\hat{a} \pm \hat{b})$ under the resonant condition $\omega_{e_1 g_1}=\omega_c$, we have obtained two hybrid modes with frequencies about $\sqrt{J}g$ departed from the optical cavity. These hybrid modes are responsible for the peaks observed, and the simple expression explains very well the change of peaks due to the varied $J$ number. This analysis indicates that the observed effect is the Rabi splitting in the strong coupling regime, and this splitting can be controlled by the optical pumping. A similar effect in the microwave domain was also discussed in our previous study~\citep{WuQ}, and confirmed in the experiment~\citep{FaheyDP}.

In Fig. \ref{fig:23models-mean}, we have further examined the mean-field results when the NV centers are treated as the three-level and two-level systems. We confirmed that both models recapture all the results qualitatively except that the latter model overestimates slightly the radiation because the NV centers trapped in the singlet excited levels also contribute to the coupling with the cavity in this model.

\subsection{Ensemble Evolution in Dicke States Space\label{sec:Dicke}}

The above studies indicate that the model treating the NV centers as two-level systems can capture most of the collective effects. By solving this model with the density matrix within the Dicke state space, it becomes possible to obtain more insights by examining simultaneously the Dicke states population and the $g^{(2)}(\tau)$-function (Fig.~\ref{fig:Dicke}). Here, we considered the systems with $40$ NV centers, as examined in the previous study~\citep{PallmannM}.

We observe that the population is distributed among the Dicke states along or slightly above the lower boundary for the weak and moderate pumping [Fig.~\ref{fig:Dicke}(a) and (b)], and is distributed among many states below the upper boundary for the strong pumping [Fig.~\ref{fig:Dicke}(c)]. Note that the center of the population distribution follows the similar trend as revealed in Fig.~\ref{fig:moreNVs}(c). Accompanying with this result, the $g^{(2)} (\tau)$-function shows dramatic change. For the weak pumping, the $g^{(2)} (\tau)$-function shows a sharp peak above one, and several weak peaks inside a deep anti-bunching dip [Fig.~\ref{fig:Dicke}(d)]. For the moderate pumping, the sharp peak becomes much stronger and broader, and the small peaks become also larger than one [Fig.~\ref{fig:Dicke}(e)]. For the strong pumping, the sharp bunching peak is replaced by a bunching dip at zero delay time and two bunching shoulders at later time [Fig.~\ref{fig:Dicke}(f)]. 

In the following, we relate the change of the $g^{(2)} (\tau)$ function with the population of the Dicke states. For the weak pumping, the excited Dicke states with $M>-J+1$ are slightly populated, and the states with larger $J$ couple strongly with the optical cavity. As a result, the NV centers can emit several photons simultaneously, leading to the sharp but weak bunching peak, and they can also absorb photons at later time, leading to the fast Rabi oscillations. For the moderate pumping, the states with $M>-J+1$  are strongly populated, but they couple slightly weakly with the optical cavity due to the reduced $J$. As a result, the NV centers can emit more photons simultaneously, leading to the sharp bunching peak. For the strong pumping, the NV centers occupy many the Dicke states with $M>-J$ and $M>0$ (population inversion), but they couple relatively weakly with the optical cavity due to the reduced $J$. Since the inverted emitters can generate superradiant pulse within the collective weak coupling regime, we attribute the two bunching peaks to the superradiant pulse [see Fig.~\ref{fig:pulses}(b)].

\section{Conclusions \label{sec:conclusion}}

In summary, in a recent experiment study~\citep{PallmannM}, we have demonstrated  the super-linear radiation and the fast photon bunching due to collective coupling of incoherently pumped NV centers with an optical cavity. In the current theoretical work, we carried out a systematical study on these effects with several sophisticated models, which treat the NV centers as five-, three- and two-level systems, and rely on the density matrix method and the mean-field approach to solve the corresponding quantum master equation. We have revealed that  the bunching shoulders in $g^{(2)} (\tau)$-function are due to the singlet excited levels, and the Rabi splitting in the steady-state spectrum can be actively controlled by the optical pumping. More importantly, the simplest model treating  NV centers as two-level systems is accurate enough to capture the most collective effects, and the populations on the excited Dicke states under the moderate pumping is responsible for the super-linear radiation and the sharp photon bunching. All in all, these results can not only guide the further experiments with the NV center systems, but are also relevant for the system with other solid-state color centers, such as  silicon-vacancy centers in diamond~\citep{BeckerJN}  and silicon-carbide~\citep{LohrmannA},  born-vacancy centers~\citep{QianC}, and carbon-related centers in hexagonal born-nitride~\citep{DowranM}.

\begin{acknowledgments}
YiDan Qu carried out the numerical calculations under the supervision of Yuan Zhang who developed the theories and the numerical programs. They contribute equally to the work. All authors contributed to the analyses and the writing of the manuscript. This work was supported by the National Key R$\&$D Program of China (No. 2024YFE0105200), the National Natural Science Foundation of China (No. 62027816), the Cross-disciplinary Innovative Research Group Project of Henan Province (No. 232300421004) and by the Carlsberg Foundation through the ``Semper Ardens'' Research Project QCooL. 
\end{acknowledgments}

\newpage
\newpage
\appendix
\renewcommand\thefigure{A\arabic{figure}}
\renewcommand\thetable{A\arabic{table}}
\renewcommand{\bibnumfmt}[1]{[S#1]}
\setcounter{figure}{0}

\section{Extra Results}

In this Appendix, we provide extra results to complement the discussions in the main text.

\subsection{Density Matrix Results for System with Few NV Centers}

\begin{figure}[!htp]
\begin{centering}
\includegraphics[scale=0.585]{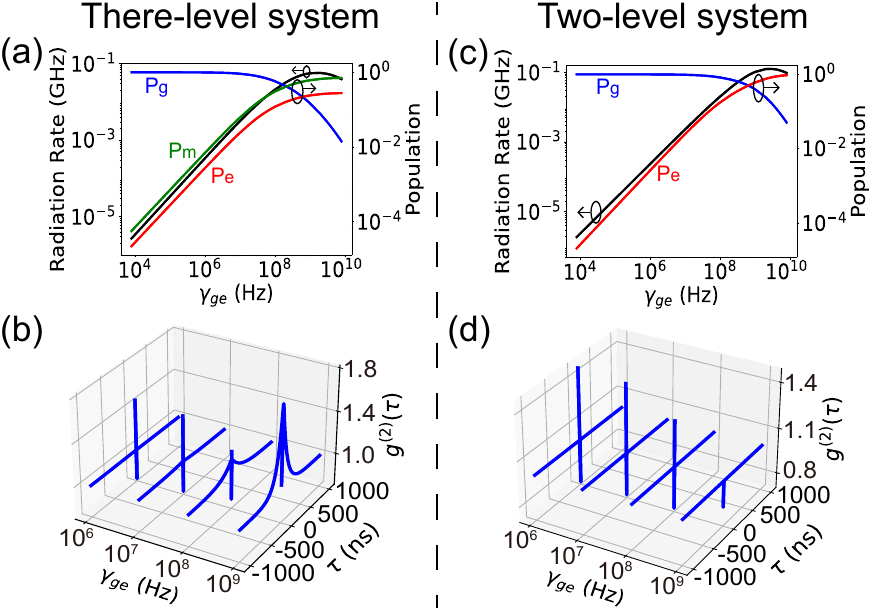}
\par\end{centering}
\caption{ \label{fig:23models}
Density matrix results based on the models treating the NV centers as three-level systems (a,b) and two-level systems (c,d). Panel (a,c) show the radiation rate (black curves, left axes), and the population of various NV levels (right axes) as function of the optical pumping rate $\gamma_{ge}$. Panel (b,d) show the $g^{(2)} (\tau)$-function as function of $\gamma_{ge}$. Here, we consider the system with two NV centers, and assume that the cavity is resonant to the NV optical transition with $m_s=0$.}
\end{figure}

In the main text, we have relied on the density matrix technique to  solve the model treating the NV centers as five-level systems, and presented the results in Fig.~\ref{fig:few}. In Fig. \ref{fig:23models}, we show the corresponding results with the models treating the NV centers as three-levels systems (a,b) and two-level systems (c,d).  In principle, the former model reproduces quantitatively the results as shown in the main text because the population on the levels related to $m_s =\pm1$ is much smaller than other levels. In contrast, the latter model can reproduce the results only qualitatively because the population on the singlet excited levels is comparable with that of the triplet excited levels. In this case, the radiation is slightly stronger since the population, which should be on the singlet excited levels, is accounted for now in the triplet excited levels, and the bunching shoulder in $g^{(2)} (\tau)$-function at longer time disappears. In any case, the latter model is accurate enough to capture qualitatively the collective effects.  

\subsection{Mean-field Results for System with Many NV Centers}

\begin{figure}[!htp]
\begin{centering}
\includegraphics[scale=0.15]{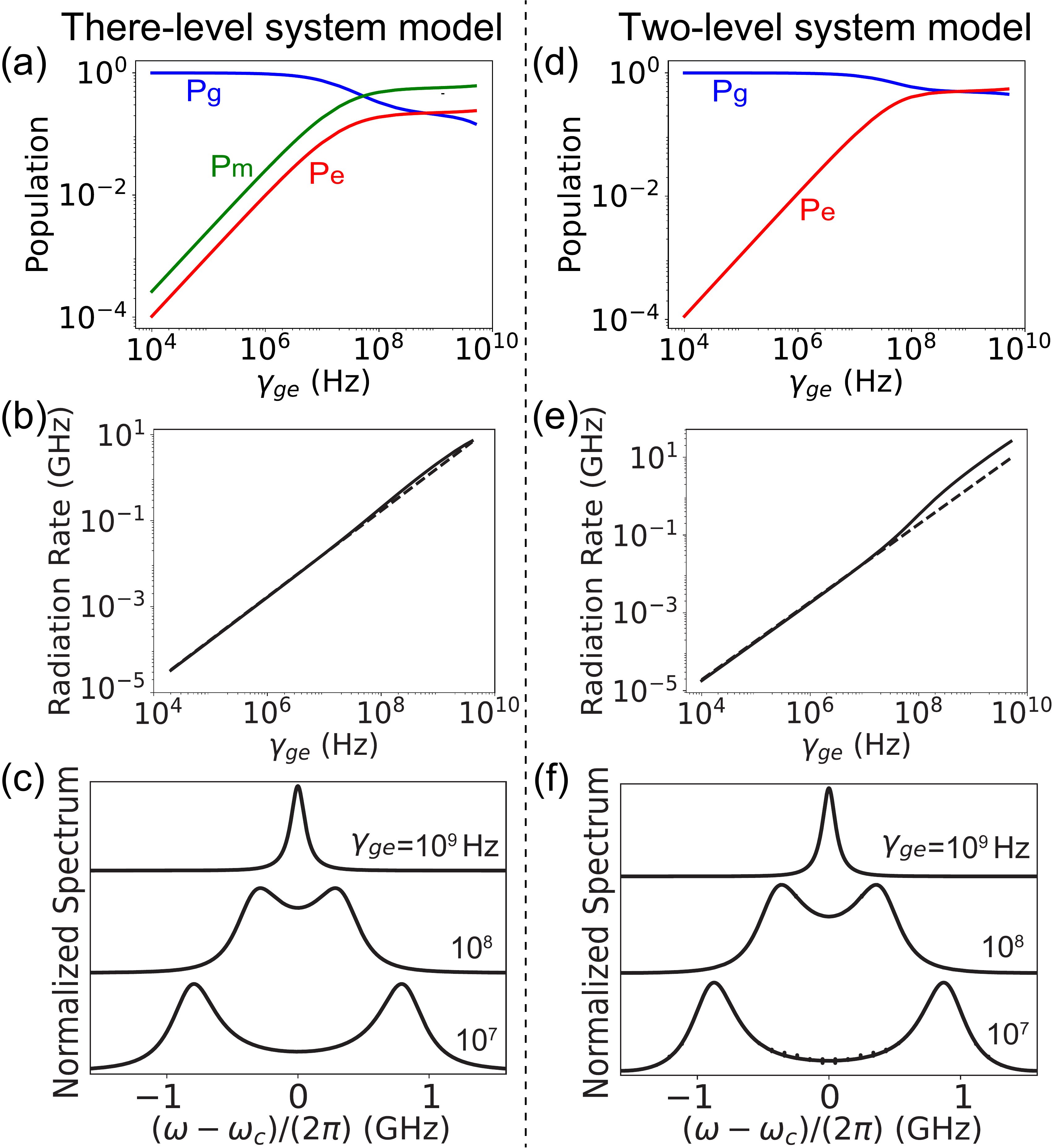}
\par\end{centering}
\caption{Mean-field results based on the models treating NV centers as the three-level systems (a-c) and two-level systems (d-f). Panel (a,d) show the populations of various NV levels as function of the pumping rate $\gamma_{ge}$. Panel (b,e) show the radiation rate as function of $\gamma_{ge}$, where the linear scaling is shown by the dashed lines. Panel (c,f) show the steady-state spectrum for three representative pumping rate $\gamma_{ge}$. Here, we consider the system with 80 NV centers.\label{fig:23models-mean}}
\end{figure}

In Fig.~\ref{fig:moreNVs} of the main text, we have adopted the mean-field approach to solve the models treating the NV centers as five-level systems. In Fig.~\ref{fig:23models-mean}, we present the complementary results for the model treating the NV centers as three-level systems (a-c) and two-level systems (d-f). Similar as the conclusions achieved in the previous subsection, the former model can reproduce quantitatively the results in the main text, while the latter model captures qualitatively most of the features except that the population on the triplet excited levels and the nonlinear radiation are overestimated slightly.

\begin{figure}
\begin{centering}
\includegraphics[scale=0.31]{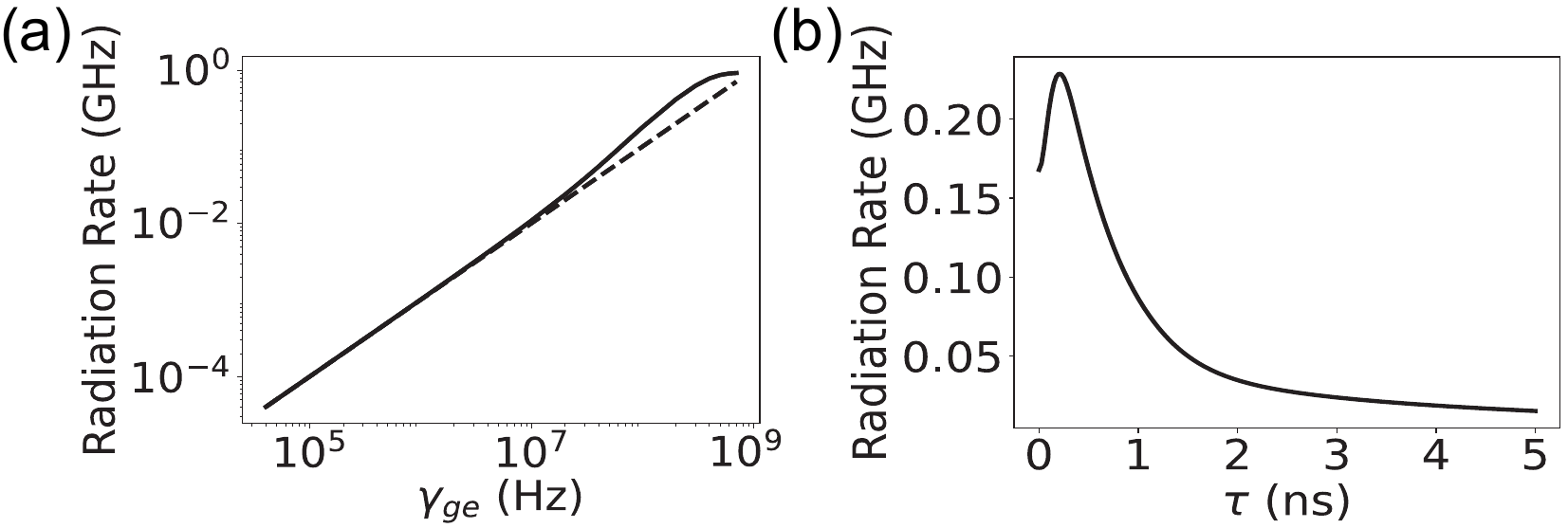}
\par\end{centering}
\caption{\label{fig:pulses} Results based on the density matrix in the Dicke states space. Panel (a) shows the steady-state radiation rate as function of the optical pumping rate $\gamma_{ge}$, where the linear scaling is shown as the dashed line. Panel (b) shows the superradiant pulse of the NV centers, which is initially excited to the Dicke states with $M>0$ by an optical pumping with rate $2\times10^8$ Hz. Here, $40$ NV centers are considered.}
\end{figure}

\subsection{Results for System in Dicke States Space}

In Fig.~\ref{fig:Dicke} of the main text, we have treated the NV centers as two-level systems, and studied the system dynamics by solving the density matrix in the Dicke states space. In Fig.~\ref{fig:pulses}, we complement these results with more details. Figure~\ref{fig:pulses}(a) shows the steady-state radiation rate as function of the optical pumping rate $\gamma_{ge}$. We observe that the non-linear scaling of radiation starts for $\gamma_{ge}$ larger than $10^7$ Hz, which can be attributed to the population of the excited Dicke states with $M>-J+1$ [Fig.~\ref{fig:Dicke}(b)].  Figure~\ref{fig:pulses}(b) shows the radiation rate pulse, i.e. the superradiant pulses, for the NV enters, which are initially prepared to the Dicke states with $M>0$  by an incoherent pumping with rate $2\times 10^8$ Hz [Fig.~\ref{fig:Dicke}(c)].

\begin{figure*}
\begin{centering}
\includegraphics[scale=0.8]{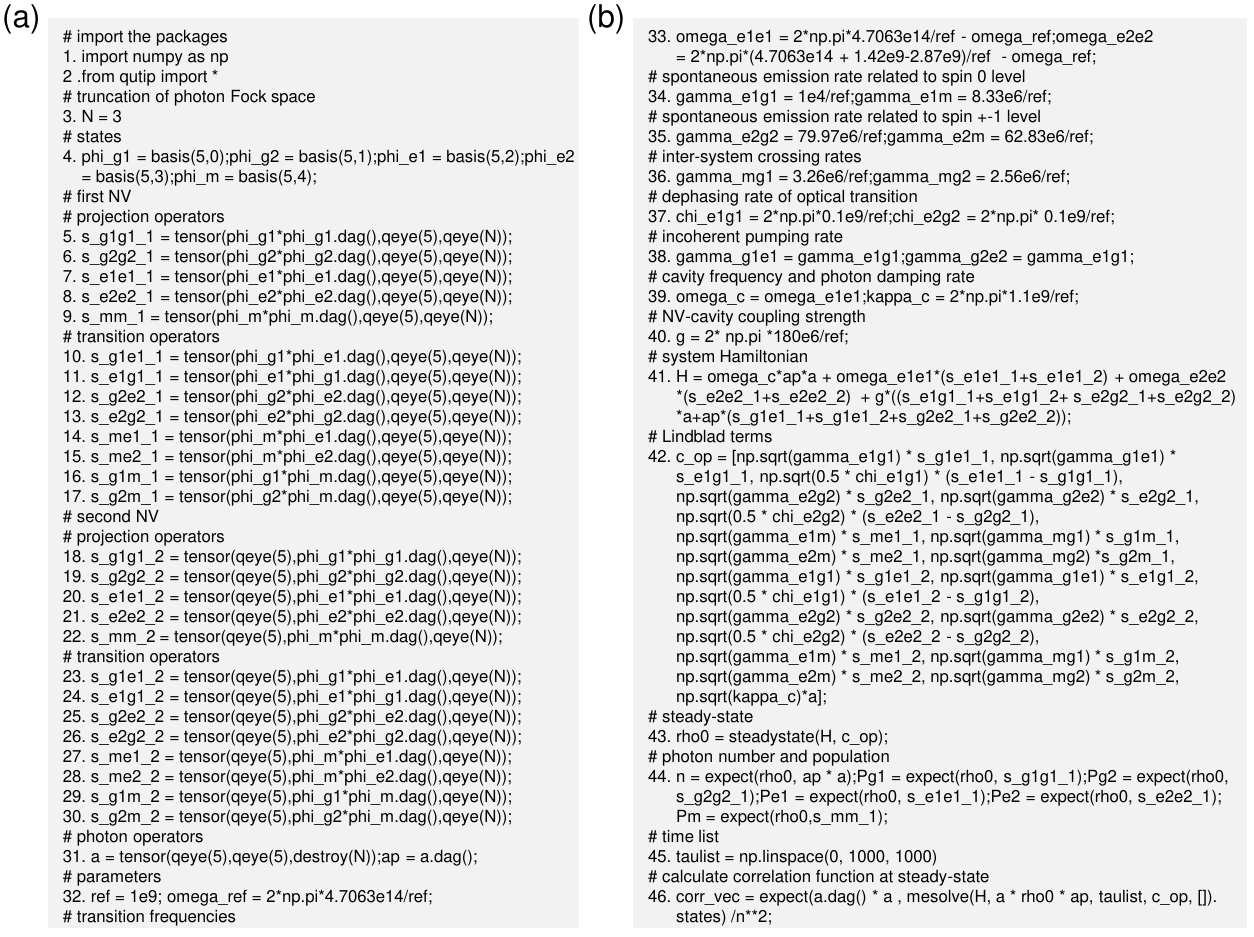}
\par\end{centering}
\caption{\label{fig:PythonCodes} Python codes to solve the QME (\ref{eq:qme}) with the standard density matrix technique as implemented in the QuTiP package.}
\end{figure*}

\section{Numerical Codes to Solve Density Matrix  Equation \label{sec:dma}}
In this Appendix, we explain how to solve QME (\ref{eq:qme}) with the density matrix technique. To deal with the model treating the NV centers as few multiple-level systems, we employ the density matrix in the product states space. To handle more efficiently the model treating the NV centers as many identical two-level systems, we utilize the density matrix in the Dicke states space.

\subsection{Codes to Solve Density Matrix Equation in Product States Space}

In the following, we explain shortly the codes  to solve the QME (\ref{eq:qme}) with the density matrix method in the product states space. First, we consider the one with the standard density matrix technique (Fig. \ref{fig:PythonCodes}). The 1st and 2nd lines import the necessary packages. The 3rd line defines the number of photon states considered. The 4th line defines the NV centers as the five-level systems. The 5th to 17th lines define the projection and transition operators, which are used latter, for the first NV center. Here, we construct the system Hilbert space as the product space of two five-level systems and optical cavity (as quantum harmonic oscillator), and define the operators as those in the corresponding sub-space. The 18th to 30th lines define the operators for the second NV center. The 31th line defines the creation and annihilation operator of the photons. The 32th to 40th lines specify the parameters of the NV centers and the optical cavity. The 41th line defines the system Hamiltonian, and the 42th line defines the list of the operators, which are used latter to define the Lindblad terms. The 43th line calculates the steady-state of the system, and the 44th line computes the mean photon number and extracts the population of NV levels at steady-state. The 45th line defines the list of time, and the 46th line calculates the auto-correlation function for the system at steady-state.

By removing the levels $g_2,e_2$ in the above model and codes, we can achieve the model treating the NV centers as three-level systems. By further removing the level $m$, we can obtain the model treating the NV centers as two-level systems. In those cases, the number of density matrix elements is reduced, and it is possible to simulate the systems with more NV centers. 

\begin{figure*}
\begin{centering}
\includegraphics[scale=1.0]{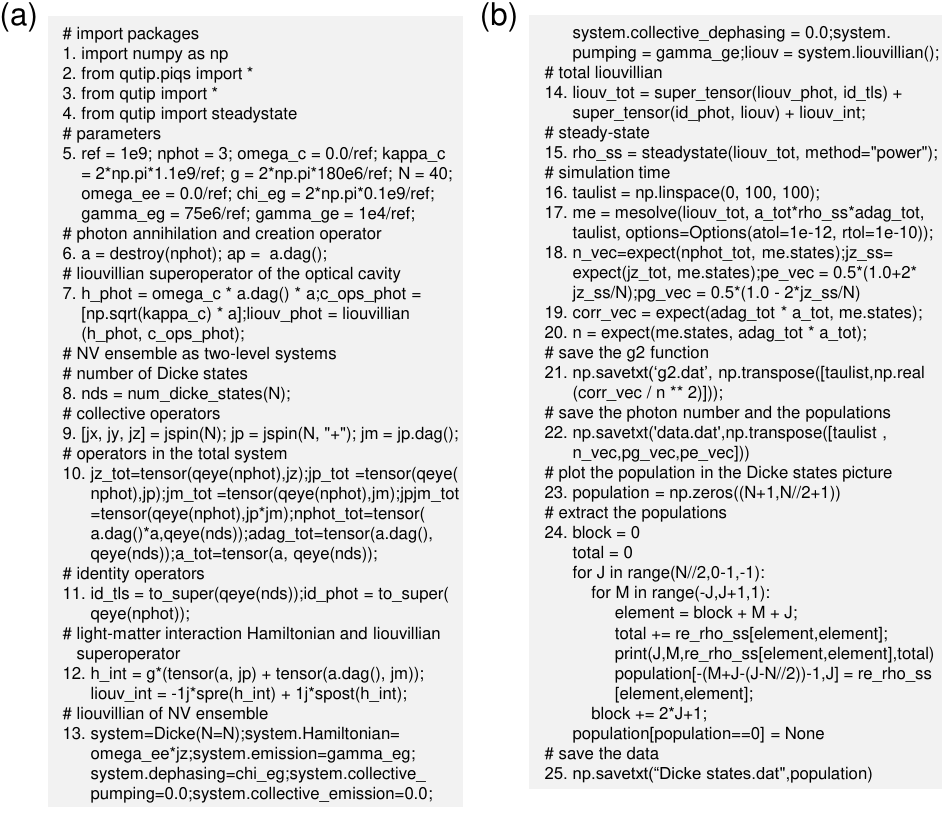}
\par\end{centering}
\caption{\label{fig:PythonCodesDicke} Python codes to solve the QME (\ref{eq:qme-tls}) with the density matrix technique within Dicke states space as implemented in the QuTip package.}
\end{figure*}

\subsection{Codes to Solve Density Matrix in Dicke States Space\label{sec:dmDicke}}

With the simplified  models, we can simulate the systems with more NV centers. However, these models can not be applied to systems with more than tens of NV centers because the Hilbert space increases exponentially with the number of NV centers. To overcome this problem, we can solve the master equation with the density matrix technique in Dicke states space.  In Fig.~\ref{fig:PythonCodesDicke}, we present the corresponding codes.  Here, the 1st to 4th lines import the necessary packages. The 5th line specifies the parameters of the NV centers and the optical cavity. The 6th line defines the annihilation and creation operator of the photons. The 7th line defines the Liouvillian superoperator of the optical cavity. The 8th line defines the number of Dicke states, where NV ensemble are treated as two-level systems. The 9th and 10th lines define the collective operators and the operators of the total system. The 11th line defines the identity operators for the sub-systems. The 12th line defines light-matter interaction Hamiltonian and Liouvillian superoperator. The 13th line specifies the parameters related to Eq.\eqref{eq:qme-tls} and the 14th line defines the total Liouvillian superoperator. The 15th line calculates the steady-state of the system, and the 16th line defines time list of the g2-function. The 17th line defines the problem, and the 18th line computes the mean photon number and extracts the population of NV levels at steady-state. The 19th and 20th lines calculate g2 function, and the 21 and 22th lines save the results into a text file. The 23th line calculates the population in the Dicke states picture, and the 24 line extracts the population. The 25th line saves the population into a text file.

\begin{figure*}
\begin{centering}
\includegraphics[scale=0.8]{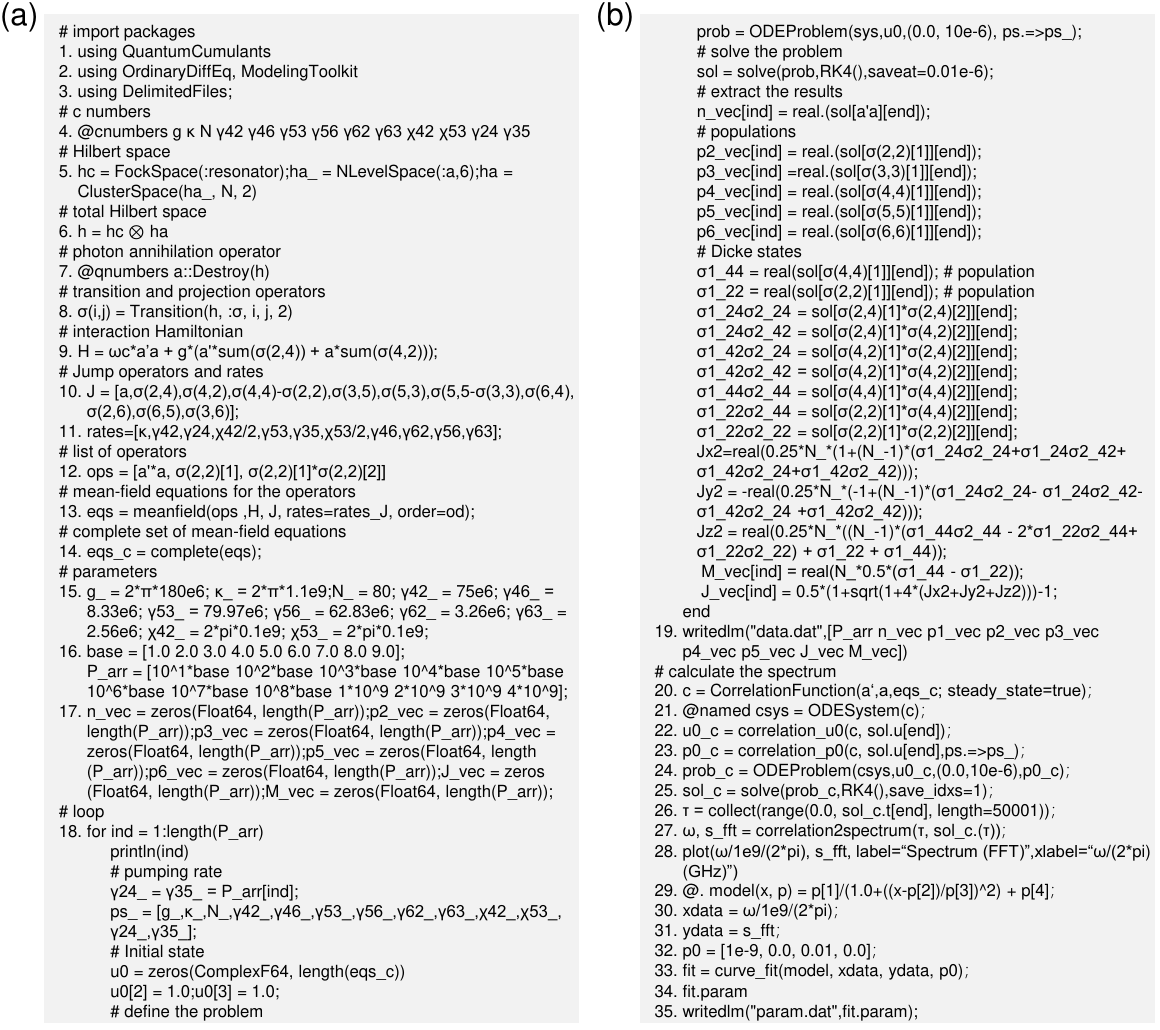}
\par\end{centering}
\caption{\label{fig:JuliaCodes} Julia codes to solve  the QME (\ref{eq:qme}) and calculate the steady-state spectrum with the mean-field approach as implemented in the QuantumCumulants.jl package.}
\end{figure*}

\section{Codes to Solve Mean-field Equations \label{sec:mf}}

In this Appendix, we present the codes to solve the QME~\eqref{eq:qme} with the cumulant mean-field approach. First, we label the involved levels as $g_{1}\to2,g_{2}\to3,e_{1}\to4,e_{2}\to5,m\to6$ to meet the convention of the QuantumCumulants.jl package~\citep{PlankensteinerD}. Here, we label the lowest level as the $2$-level, and specify the initial population on this level. If we label the lowest level as the $1$-level, we can not access directly the information related to the projection operator of the $1$-level $\hat{\sigma}_i^{11}$, which is replaced by the relation $1-\sum_{l\neq 1} \hat{\sigma}_i^{ll}$ in the QuantumCumulants.jl package to reduce the number of independent elements, and leads to  the difficulty of calculating the average of the Dicke state quantum numbers. The corresponding codes are presented in Fig. \ref{fig:JuliaCodes}.   

We now turn the codes with the mean-field approach (Fig. \ref{fig:JuliaCodes}). The 1st to 3rd lines import the necessary packages. The 4th line defines the complex parameters. The 5th line defines the Hilbert space for the cavity as a quantum harmonic oscillator, the NV centers as six-level systems, and the 6th line defines the product of these spaces as the one for the system. In the QuantumCumulants.jl package, the first level does not appear directly, but is represented by other levels through the completeness relation. Thus, to better extract the information on the first level, we consider six levels to represent the five-level system by leaving  the first level unused. The 7th line defines the annihilation operator of the cavity photons, and the 8th line defines the transition and projection operators of the NV centers. The 9th line defines the system Hamiltonian in interaction picture, and the 10th and 11th lines define the list of operators and rates, required to define the Lindblad terms. The 12th line defines the list of operators, whose mean-field equations are derived in the 13th line. The 14th line analyzes the unknown mean fields, and derives the equations for them. The 15th line specifies the values of the parameters. The 16th line defines the pumping rate, and the 17th line defines the vectors to store the mean fields of interest for a loop. The 18th line computes the mean-field for given pumping rate. The 19th line save the results into a text file.

The 20th to 35th lines calculate the correlation function. Here, the 20th line defines the correlation function, the 21th line defines the ordinary differential equations (ODE) system, the 22th and 23th lines compute the initial values for the correlation function and the parameters involved. The 24th and 25th lines define and solve the ordinary differential equations (ODE) problem. The 26th line defines an equidistant list of times for FFT, the 27th line computes the spectrum from the solution, the 28th line plots the results. The 29th to 34th lines fit the spectrum obtained above. The 35th line saves the fit params into a text file.

\section{Evaluation of Spectrum and g2 Function \label{sec:speccorr}}

In Sec.~\ref{sec:theory} of the main text, we have explained the expressions to compute the $g^{(2)}(\tau)$-function and the steady-state spectrum. In these expressions, the expectation values  are defined with the time-dependent operators in the Heisenberg picture, but can be transformed as those with the time-dependent density operator in the Schr{\"o}dinger picture, $\left\langle\hat{a}^{\dagger}(t_{ss})\hat{a}(t_{ss})\right\rangle=\mathrm{tr}\left\{\hat{a}^{\dagger}\hat{a}\hat{\rho}\left(t_{ss}\right)\right\} $, $\left\langle \hat{a}^{\dagger}(t_{ss})\hat{a}^{\dagger}(t_{ss}+\tau)\hat{a}(t_{ss}+\tau)\hat{a}(t_{ss})\right\rangle=\mathrm{tr}\left\{\hat{a}^{\dagger}\hat{a}\hat{\mathcal{U}}\left(t_{ss}+\tau\right)\left[\hat{a}\hat{\rho}\left(t_{ss}\right)\hat{a}^{\dagger}\right]\right\} $, where $\hat{\rho}\left(t_{ss}\right)$ denotes the reduced density operator at the steady-state, and the superoperator $\hat{\mathcal{U}}\left(t\right)$ indicates the formal solution of Eq. (\ref{eq:qme}). In practice, we evolve firstly the reduced density operator, and then compute the combined
operator $\hat{\varrho}\left(\tau=0\right)=\hat{a}\hat{\rho}\left(t_{ss}\right)\hat{a}^{\dagger}$, and finally evolve $\hat{\varrho}$  further with Eq. (\ref{eq:qme}) according
to quantum regression theorem~\citep{CGardiner2004}. The 20th to 27th line of Fig.~\ref{fig:JuliaCodes} show the codes to calculate the steady-state spectrum with the mean-field approach.

\end{document}